\documentclass{llncs}
\usepackage[utf8]{inputenc}
\usepackage{comment}
\usepackage{amsmath}
\usepackage{url}
\usepackage{blindtext}
\usepackage[dvipsnames]{xcolor}
\newcommand{\N}{\textcolor{Green}{N}\color{black}} 
\newcommand{\Y}{\textcolor{red}{S}\color{black}} 
\usepackage{array}
\newcommand{\PreserveBackslash}[1]{\let\temp=\\#1\let\\=\temp}
\newcolumntype{C}[1]{>{\PreserveBackslash\centering}p{#1}}
\newcolumntype{R}[1]{>{\PreserveBackslash\raggedleft}p{#1}}
\newcolumntype{L}[1]{>{\PreserveBackslash\raggedright}p{#1}}
\usepackage{rotating} %rotate
\newcommand*\rot{\rotatebox{90}} %rotate text 90degrees

\usepackage{amsthm}
\theoremstyle{definition}
\newtheorem{defn}{Definition}
\newcommand{\ket}[1]{\ensuremath{\left|#1\right\rangle}} % Dirac Kets
\title{Quantum multi-factor authentication}
\author{Hazel Murray\inst{1}\orcidID{0000-0002-5349-4011} \and David Malone\inst{2}\orcidID{0000-0002-6947-586X}}
\authorrunning{H.~Murray and D.~Malone}
\institute{Munster Technological University, Ireland \email{Hazel.Murray@mtu.ie} \and
Maynooth University, Ireland \email{David.Malone@mu.ie}}
\date{May 2021}

\begin{document}

\maketitle
\begin{abstract}
We present a quantum multi-factor authentication mechanism based on the hidden-matching quantum communication complexity problem. It offers step-up graded authentication for users via a quantum token. In this paper, we outline the protocol, demonstrate that it can be used in a largely classical setting, explain how it can be implemented in SASL, and discuss  arising security features. We also offer a comparison between our mechanism and current state-of-the-art multi-factor authentication mechanisms.
\end{abstract}

%Additional Questions to maybe answer: 
%1. How much security would we get if we limited it to bit strings that did not require entanglement?
%2. Look at the trade off between Gavinsky's k and t, and see if this trade-off would need to be the same.

\section{Introduction}
Multi-factor authentication (MFA) is of increasing importance  for the security of individual accounts and infrastructures. However, many multi-factor mechanisms come with security or usability drawbacks. For example, though widely used, using SMS to send security codes is an insecure form of MFA since phone numbers are easily clone-able and SMS messages may be redirected~\cite{schneiern-sms}. In addition, in the near future, some authentication mechanisms using public-key cryptography, such as certificates and some hardware devices, may face challenges from quantum computing.

In this work, we suggest a quantum multi-factor authentication mechanism based on an established quantum scheme suggested for use in the banking sector by Gavinsky~\cite{gavinsky2012quantum}. The proposed mechanism offers advantages over classical multi-factor authentication schemes by using quantum principles in order to protect against duplication and eavesdropping attacks. It also offers an ingrained ability to offer high and low levels of assurance with the same token, based on the current needs, trust and actions of the verifier and user. Benefits in comparison to other quantum mechanisms include the fact that there is no need for a quantum communication channel for each authentication, the verifier only needs to store classical strings rather than quantum registers, the token can be reused for multiple authentications of the same user, and a secure classical channel is not a hard requirement.

%Section~\ref{} compared it to a time-based one time password. Note that tokens can expire.
%Section~\ref{] shows how it can be incorporated into SASL.
%(We could show a simulation, but it might be nicer to keep it as a protocol paper). 

%In the section below we give a brief summary of the mechanism and mention the key features of note. 

\subsection{Brief summary of mechanism}\label{sec:summary}
%The proposed mechanism involves a quantum MFA token which is issued by an organisation to a user. It uses the quantum communication complexity problem known as the Hidden Matching Problem~\cite{bar2004exponential} to allow the organisation to verify that the user Alice holds the token each time she wishes to authenticate. 
The proposed mechanism involves a quantum MFA token which is issued by an organisation to a user. This token contains $k$ quantum registers which correspond to a list of $k$ classical strings held by the organisation. The relationship between the classical and quantum bits is defined by an established quantum communication complexity problem known as the Hidden Matching Problem (HMP)~\cite{bar2004exponential}.
Communication complexity describes the problem where two entities each hold information, one holds data $x$ and the other holds data $y$. They wish to perform some computation using the two pieces of information but neither wishes to reveal their data to the other. In our case, Alice holds a quantum token which contains $k$ quantum registers, and the verifier holds the $k$ corresponding classical strings. Without either revealing their data to the other, the verifier can verify that Alice holds the correct token.

Informally, the authentication steps in the protocol involve: 1. the verifier and Alice agreeing which of the $k$ registers should be measured. 2. The verifier then tells Alice which basis should be used for the measurements. 3. Alice will return a pair of values $(a,b)$ to the organisation which correspond to the outcome of each measurement. 4. Using the results from Alice, the stored classical strings, and knowledge of what basis was used, the Hidden Matching Problem allows the organisation to verify that Alice does indeed hold the token. 

%This problem means that by knowing the classical bits, the organisation can ask for measurements of the users' quantum registers and will be able to verify that the user holds the correct token. Gavinsky's 2012 paper offers formal proofs of the security of this verification mechanism~\cite{gavinsky2012quantum}.  Below we list the key security and utility features: %We believe these mechanisms show promise as a method of multi-factor authentication. There are a number of benefits of the mechanism we propose. These are:%Firstly, quantum bits obey the ``no cloning theorem'' meaning that many cloning, duplication or replay attacks for compromising authentication can be lessened. 
%Second, the mechanism takes the form of a zero-knowledge protocol and therefore a secure communication channel is not a hard-requirement for the authentication.
Below we list some of the key security and utility features:
\begin{itemize}
    \item The level of assurance depends on the number of measurements the verifier requests. This means that a single token has an  ingrained  ability  to  offer  high  and  low  levels  of  assurance based on the current needs, trust and actions of both verifier and user. This is valuable in a multi-factor mechanism, where a goal is to increase the assurance provided by a first factor.
    \item When a quantum register is selected for measurement it can not be used again. This means that there is a trade-off between the lifetime of the token and the level of assurance requested each time. This natural lifetime of an authentication token can be a valuable security feature. It also means that measurement results from a previous authentication have limited value to an attacker and therefore a secure communication channel is not a hard-requirement for the authentication.
    \item The quantum registers chosen for measurement are mutually agreed between  verifier and  user, so neither has complete control over the registers used. 
    \item Despite being a quantum protocol, the authentication exchange does not require a quantum communication channel between  user and  verifier as only the measurement requests and results need to be communicated between them. This makes it viable for integration in largely classical settings.
    \item Also, in this mechanism the organisation only needs to hold classical strings rather than quantum registers. This is an important improvement, as holding quantum registers corresponding to each user who wants to authenticate would lead to significant overhead for an organisation.
    \item The security of the scheme is based on the premise of a zero-knowledge proof, where the user can prove that they hold the token without revealing the token. Unlike other schemes, there are no requirements for auxiliary classical cryptography. 
    \item However, it is important to note that, at the time of writing, the short length of time quantum memories can hold qubits means a true hardware implementation is not currently possible.
\end{itemize}

%This is nicer wording %Thirdly, the mechanism can provide ``degrees of security'' depending on the level of risk or trustworthiness of the actor or action. Though work has been done to link risk to the level of authentication required, so far this has not been embedded within multi-factor authentication mechanisms. 
%Related to this feature, the lifetime of the token is limited based on its designed size. In situations where indefinite access to a system is not recommended, this in-built feature could be beneficial to the securer.
%Section~\ref{sec:mechanism} provides a detailed description of the mechanism and protocol and Table~\ref{tab:compare} gives a visual comparison of our mechanism to the current state-of-the-art. 

The remainder of the paper is organised as follows. In Section~\ref{sec:related} we give background on the development of quantum authentication protocols to date. 
In Section~\ref{sec:background-quantum}, we present the quantum preliminaries necessary for the understanding of the mechanism. 
In Section~\ref{sec:mechanism}, we present the quantum multi-factor authentication protocol and the associated SASL mechanism. This section also includes an informal security analysis and explanation and discussion about the token lifetime. In Section~\ref{sec:state-art}, we discuss the proposed mechanism in relation to current state-of-the-art multi-factor authentication. Section~\ref{sec:disc} includes a discussion and we conclude in Section~\ref{sec:conc}

%Include the related work here. 
\section{A brief review of quantum authentication}\label{sec:related}
As with classic authentication, a distinction exists between message authentication and identity authentication. In this paper, we are concerned with identity authentication, that is, Alice authenticates her identity to the server, usually using a pre-agreed secret. For example, in authentication using passwords, the password for Alice is agreed at the beginning of the communication exchanges, and then for each subsequent authentication, the user provides the password. Quantum authentication protocols will typically use quantum states as the pre-agreed secret and a user can then authenticate using these, usually through some type of measurement strategy. We will now describe the development of some of these quantum identity authentication mechanisms that have been proposed.

Quantum authentication protocols can be roughly divided up into four kinds: those based on entangled pairs, those based on a quantum key distribution (QKD), those based on superposition, and those based on a quantum computation complexity problem. To the best of our knowledge, ours is the first based on the last option. Here we will discuss the other three with reference to their positives and negatives. 

\paragraph{Quantum authentication via entangled qubits}
In 2000, Zeng and Guo~\cite{zeng2000quantum} introduced an identity authentication quantum protocol based on symmetric cryptography and EPR pairs (maximally entangled quantum states of two qubits) which have been previously shared. In their protocol, there must exist a pre-agreed key $K_1$. They will use this key to decide which basis to use for measurements. When Alice wants
to secretly communicate with Bob, Alice and Bob set up a quantum channel which consists of a source that emits EPR pairs. Alice receives one half of each entangled pair and Bob receives the other. Alice performs a series of measurements (according to their key) on her half of the EPR pairs. Bob, in his turn, measures his half with the same key and also performs a random series of measurements $M$. If
eavesdropping occurred, Alice and Bob can detect it using the random series of measurements, and then can identify each other by comparing the measurements done using their shared  key. These results are exchanged via classical symmetric key cryptography. As it still requires classical cryptography and a secure classical channel it is not necessarily secure from quantum computing attacks.

This scheme was preceded by a similar scheme by Barnum~\cite{barnum1999quantum}. It was also developed on by Li and Zhang in 2004~\cite{li2006quantum}, Li and Barnum in 2006~\cite{li2004quantum} and Ghilen et al. in 2013~\cite{ghilen2013quantum}. All these schemes require quantum communication channels. Li and Barnum's protocol requires no previous key to be shared between the parties, just entangled
qubits. It also does not require any classical communication but does still need quantum communication of the qubits at each authentication. Ghilen et al.'s protocol~\cite{ghilen2013quantum} allows the state $\phi^+$ to be represented as
binary bit ``0'' and the state $\psi^-$ to be represented as binary bit ``1'', and thus includes key agreement as part of the protocol.

In 2020, Sharma and Nene~\cite{sharma2020two} proposed an entanglement-based two factor authentication scheme which combines the measurement of entangled qubits with a biometric-based secret to achieve authentication. We will discuss this scheme in more detail in Section~\ref{sec:state-art} where we compare our mechanism to the current multi-factor state-of-the-art.

%"Barnum [10] proposes a quantum identity authentication protocol that exploits the phenomenon of entanglement-catalyzed transformations between pure states. Alice and Bob share a catalyst state |y>, and there are incommensurate states |v1> and |v2> such that in the presence of the catalyst, |v1>  can be converted to |v2>  while retaining |y> . When Alice wants to authenticate, Bobprepares |v1>  and sends half of it to her. They go through the steps, involving local measurements, one-way communication of measurement results, and local operations conditional on those measurements results, which convert |v1>  to |v2> . This protocol involves qubits exchanges and classical communication"

%"The protocol proposed by Zeng and Zhang [11] uses a trusted center to help the legitimate users to authenticate identity. The trusted center sets up a quantum channel between Alice and the center and between Bob and the center. The center generates the same two entangled pairs to Alice and Bob, keeping half of each. Similarly to BB84, Alice and Bob measure their particles with a randomly chosen basis (horizontal-vertical or diagonally polarized) and share the basis used for the measurements, creating a session key – so, in this protocol, both authentication and QKD are implemented."

\paragraph{Quantum authentication via  QKD and classical cryptography}
Quantum key distribution is the most established form of quantum cryptography. It allows two parties to mutually agree a security key. The most famous example is the BB84 protocol~\cite{bennett1983quantum} which is widely deployed. %The protocol is based on three quantum properties: 1. The no-cloning theorem, which makes quantum states impossible to duplicate, something that hinders an eavesdropper (Eve) from intercepting the quantum communications, 2. Measurements lead to the collapse of the quantum states. 3. Measurements are non reversible. BB84 requires both a quantum and a secure classical communication channel.
In this protocol Alice begins by sending polarized photons, set using one of two bases, to Bob. Bob observes the received photons and randomly chooses which basis he will measure with respect to. Alice and Bob then use a classical channel to determine which data bits should be discarded by exchanging information about the bases they used for the measurements. They can now use the results which were measured using the same basis as their shared key. 

Du{\v{s}}ek~\cite{duvsek1999quantum} proposed an authentication scheme where the BB84 QKD is used to share an identification sequence. After Alice and Bob share these secret codes, they use a classical channel. They send parts of the identification sequence to each other to demonstrate that they have it. However, an additional authentication is required because the BB84 needs authentication before the parties start
communicating.

Kuhn~\cite{kuhn2003hybrid} proposed a new authentication scheme which used both QKD and classical cryptography. It requires a trusted server who holds a shared secret key with both Alice and Bob. If Alice wants to authenticate Bob, this protocol can then be used. The trusted server sends a stream of authentication
bits via QKD on a quantum channel to Alice and Bob, such that Alice and Bob each get one part of a pair of entangled qubits. On a secure classical channel, the trusted server tells them which basis to measure each bit with respect to. Alice can then send the results of a portion of her measurements to Bob to authenticate herself. The remaining bits are kept and used as a shared secret key. A positive feature of this protocol is that the trusted server does not learn the shared secret key between Alice and Bob that is used for their future communication. 

\paragraph{Quantum authentication via superposition}
Quantum authentication protocols which rely on the sharing of entangled qubits have practical drawbacks. In particular, they are not very scalable, as a verifier would need to maintain qubits in a superposition state for each user who needs to authenticate. In response to this, quantum authentication schemes which rely on superposition rather than entanglement were put forward as a solution. 

In 2017, ho Hong et al.~\cite{ho2017quantum} proposed an authentication protocol which uses single photon states. The two parties, Alice and Bob, have a pre-agreed secret string. If Alice wishes to authenticate to Bob, she encodes the classical secret string into corresponding quantum registers. She then sends these to Bob. Bob can measure these and, if the output matches the stored string corresponding to Alice, then Alice is authenticated. The authors compare this scheme to the verification of a password using a password hash.  
This protocol offers good efficiency, however, errors in the transmission or generation of the photons will mean this mechanism will fail. 
In 2019, Zawadzki~\cite{zawadzki2019quantum} addressed the information loss to eavesdroppers problems which also existed in the protocol. However, in 2021, Gonz{\'a}lez-Guill{\'e}n et al.~\cite{gonzalez2021attack} identified an attack on Zawadzki's algorithm and demonstrated that it was insecure.

\paragraph{Quantum authentication via communication complexity}
As mentioned, the scheme we propose for multi-factor authentication (inspired by Gavinsky~\cite{gavinsky2012quantum}) is based on the quantum communication complexity problem known as the Hidden Matching Problem~\cite{bar2004exponential}. The organisation is able to issue Alice with a token which she can then use for future authentication. Alice never needs to pass quantum bits but instead sends to the results of certain measurements as they are requested by the organisation. The organisation is able to verify that Alice holds the correct token based on these responses. It takes the form similar to a zero-knowledge proof and therefore and eavesdropper learns very little from multiple observations of the protocol. %Communication complexity describes the problem where two entities each hold information, one holds data $x$ and the other holds data $y$. They wish to preform some computation using the two pieces of information but neither wishes to reveal their data to the other. In our case, Alice holds a quantum token, and the verifier holds corresponding classical strings. Without either revealing their data to the other, the verifier can verify that Alice holds the correct token.

%\paragraph{Our quantum authentication protocol via quantum communication complexity}
%Our protocol involves the verifier sending a quantum token to the user, Alice. This quantum token contains $k$ two-qubit registers which correspond to a set of classical strings which the verifier holds on to. When Alice wishes to authenticate, they will mutually agree a set of registers which can be measured. Alice then returns the outcome of these measurements, and using only the stored classical strings, the verifier will be able to verify with high probability that Alice does in fact hold the corresponding token. %The token can be used for multiple authentications. 

One important distinction between our scheme and the above schemes is that the verifier does not need to hold the quantum states, which can become unmanageable for a large number of users. The verifier, in our scheme, need only store classical bits which can be reused for multiple authentications of the same user. A second distinction is the ability to offer differing levels of security within the scheme. This is particularly useful for application to multi-factor authentication (MFA), as MFA is often used as step-up authentication when risky actions are attempted by a user.

\section{Quantum computing properties and preliminaries}\label{sec:background-quantum}
Quantum computing has the theoretical power to break certain modern cryptography \cite{mavroeidis2018impact}. In 1994, Peter Shor~\cite{shor1994algorithms}, developed a quantum algorithm which threatens public key cryptographic systems which are often used in multi-factor authentication devices~\cite{kunnemann2012yubisecure}. In 1996, Grover's algorithm was developed, which reduced the effectiveness of symmetric key cryptographic systems \cite{grover1996algorithms}. Without these classical cryptographic mechanisms, many of our secure authentication protocols become vulnerable. 

Though quantum computing has revealed weaknesses in current cryptographic mechanisms, it also holds the possibility of unlocking solutions that exceed the bounds of our current computational capabilities. Quantum mechanical systems have properties that are at odds with our general understanding of classical physics and here we will give a brief overview of the properties we utilise.

\paragraph{Qubit}
In classical computing, all computation is done using bits which can be either 0 or 1. A qubit is the quantum equivalent to this classical bit but has a number of important properties. These are included below. 
 
\paragraph{Superposition}
The first property is \textit{superposition}. This describes the fact that a qubit can take the value of 0 and 1 or both at the same time. This gives quantum computers the capacity, in some sense, to complete computations in parallel and where $n$ classical bits allow $n$ computations, $n$ qubits can allow $2^n$ computations.

\paragraph{Measurement}
The second property is \textit{measurement}. In classical mechanics, looking at something does not change its state. In quantum mechanics, a qubit can be in a \textit{superposition} of both 0 and 1 at the same time and when measured it must \textit{collapse} to either 0 or 1. The state of a quantum bit is represented by a wave function, where $\ket{0}$ is the 0 wave function, $\ket{1}$ is the 1 wave function, and $\alpha\ket{0}+\beta\ket{1}$ is a superposition. If we measure a wave function then there is a probability $|\alpha|^2$ of measuring 0 and $|\beta|^2$ of 1. Therefore, the result we get is probabilistic and not predetermined. 

\paragraph{Basis}
When we take measurements, we measure a particular property of the qubit. Similar to how measuring speed and weight of a classical object will result in different results, so too will measuring a qubit with respect to different \textit{bases}. We take advantage of this and it allows the mechanism to take a form similar to a zero-knowledge protocol since measurement with respect to one basis will give no information regarding the correct measurement when done with an orthogonal basis. This means that we can offer security for the mechanism even when a secure communication channel is not available. In this paper, we assume a qubit can be measured using two particular \textit{bases}. %xxx

\paragraph{Entanglement}
According to Einstein, \textit{entanglement} is the `spooky' property of quantum mechanics. Let's say we \textit{entangle} two qubits and move them to opposite sides of the globe. If we measure one of the qubits then we know that we will get the same measurement for the second, entangled, qubit. Imagine we take the first qubit and measure it using a momentum basis and get 1. Then the other qubit will also measure as 1. This is remarkable since each returned result is a function of probabilities $|\alpha|^2$ and $|\beta|^2$. 

\paragraph{No-cloning theorem}
The final property we will mention is the \textit{no-cloning theorem}~\cite{wooters1982quantum} which tells us that it is impossible to create an identical copy of a quantum state. This has particularly relevant applications for one-time-passwords~\cite{sharma2020two} and secure authentication. In our mechanism, this means that the risk of cloning, duplication or replay attacks is significantly reduced.

\subsection{Hidden Matching Problem} 
Our Quantum MFA mechanism is based on the quantum-classical communication complexity problem known as the Hidden Matching Problem~\cite{bar2004exponential}. The specific version of the problem known as $\textit{HMP}_4$ sets out a relationship between a 4-bit classical string and a 2-qubit quantum register (called the $\textit{HMP}_4$-state). By requesting measurements of these states in a particular format (the $\textit{HMP}_4$-reply), the verifier can conclude whether the user holds the correct token (this is the $\textit{HMP}_4$-condition).  In this section, we will describe these three key aspects of $\textit{HMP}_4$ in more detail. 

We begin by explaining how the quantum registers should be created so that they correspond to the classical bit strings held by the organisation ($\textit{HMP}_4$-state). Here, a classical string is denoted $x$ and the quantum register corresponding to that classical string is denoted $\ket{\alpha(x)}$. %The quantum bits are created so that they satisfy the following form: 

\begin{defn}[$\textit{HMP}_4$-states]\label{def:HMP-states}
    Let $x \in \{0,1\}^4$. The corresponding quantum register is 
    \[
        \ket{\alpha(x)} = \frac{1}{\sqrt{4}} \sum_{1 \leq i \leq 4} (-1)^{x_i} \ket{(i-1)_2},
    \]
    where $(\cdot)_2$ denotes writing a number in base 2.
\end{defn}
\noindent In 2020, Murray et al. showed how qubit registers of this form can be created using quantum gates~\cite{murray2020implementing}. 

Once these quantum registers are created they can be passed to the user. At each authentication, the user will measure a selection of these registers and the organisation will use the value $m$ to will tell the user which basis to use. The user does not return the results directly but instead returns an $(a,b)$ pair based on the following rule: 
\begin{defn}[$\textit{HMP}_4$-reply]
 \label{def:HMP-queries} \hphantom{hello}
   
\noindent If $m = 0$, 

    \begin{align*}
        v_1 = \frac{\ket{00} + \ket{01}}{\sqrt{2}}, v_2 = \frac{\ket{00} - \ket{01}}{\sqrt{2}}, v_3 = \frac{\ket{10} + \ket{11}}{\sqrt{2}}, v_4 = \frac{\ket{10} - \ket{11}}{\sqrt{2}}
    \end{align*}
\noindent    Otherwise if $m=1$,

    \begin{align*}
        v_1 = \frac{\ket{00} + \ket{10}}{\sqrt{2}}, v_2 = \frac{\ket{00} - \ket{10}}{\sqrt{2}}, v_3 = \frac{\ket{01} + \ket{11}}{\sqrt{2}}, v_4 = \frac{\ket{01} - \ket{11}}{\sqrt{2}}
    \end{align*}
    Measure \ket{\alpha(x)} in the basis $\{v_1,v_2,v_3,v_4\}$. Return $(a,b)$ such that: %and let $(a,b)$ be $(0,0)$ if the outcome is $v_1$; $(0,1)$ in the case of $v_2$; $(1,0)$ in the case of $v_3$; $(1,1)$ in the case of $v_4$. 
 \[
    (a,b)= 
\begin{cases}
    (0,0) ,& \text{if } v_1\\
    (0,1) ,& \text{if } v_2\\
    (1,0) ,& \text{if } v_3\\
    (1,1) ,& \text{if } v_4.\\
\end{cases}
\]
\end{defn}

Using the initial classical strings that the organisation stores, $x$, the basis indicator, $m$, and the reply from the user, $(a,b)$, the verifier is now able to validate whether these values satisfy the $\textit{HMP}_4$-condition. For each measured register, the verifier checks that:
\begin{defn}[$\textit{HMP}_4$  condition]\label{def:HMP}
For $x \in \{0,1\}^4$ and $m,a,b \in \{0,1\}$, we say that
    $(x,m,a,b) \in \textit{HMP}_4$ if \[
    b= 
\begin{cases}
    x_1 \textsc{ xor } x_{2+m} & \text{if } a= 0\\
    x_{3-m} \textsc{ xor } x_4 & \text{if } a=1
\end{cases}
\]
\end{defn}
%Note, the number of different registers that are measured corresponds to the level of assurance the verifier receives.
%Note, there is some probability ($\frac{1}{4}$ with no prior knowledge, and $\frac{1}{2}$ with prior knowledge) that an attacker might be able to correctly guess the result $(a,b)$ so the number of different registers that are measured correspond to the level of assurance the verifier receives. %check probs xxx

\section{Quantum MFA mechanism}\label{sec:mechanism}
In this section we will show how these quantum properties and the $\textit{HMP}_4$ problem can be leveraged to create a secure multi-factor authentication mechanism. The design of the scheme is closely based on Gavinsky's quantum coin design \cite{gavinsky2012quantum} and inspired by the use of traditional hardware tokens, such as in TOTP \cite{m2011totp}.

This mechanism had two phases: \textbf{Issuing} and \textbf{Authentication}. These are explained separately below. Note that once a token has been issued it can be used for authentication multiple times.

\subsection*{Issuing}
In the first phase, a quantum token is issued to a user. This part requires quantum communication, where a set of quantum states are created by the token issuer (server) and transferred into quantum memories held by the user. This could be achieved by issuing a device with quantum memories to the user. The user can then use this quantum token for authentication. The server must store the classical bit strings used in the creation of the token. However, as this information is classical, it can be stored in a secret authentication database, similar to the information stored for a classical hardware token.

\begin{figure}
    \centering
    \includegraphics[width=0.6\linewidth]{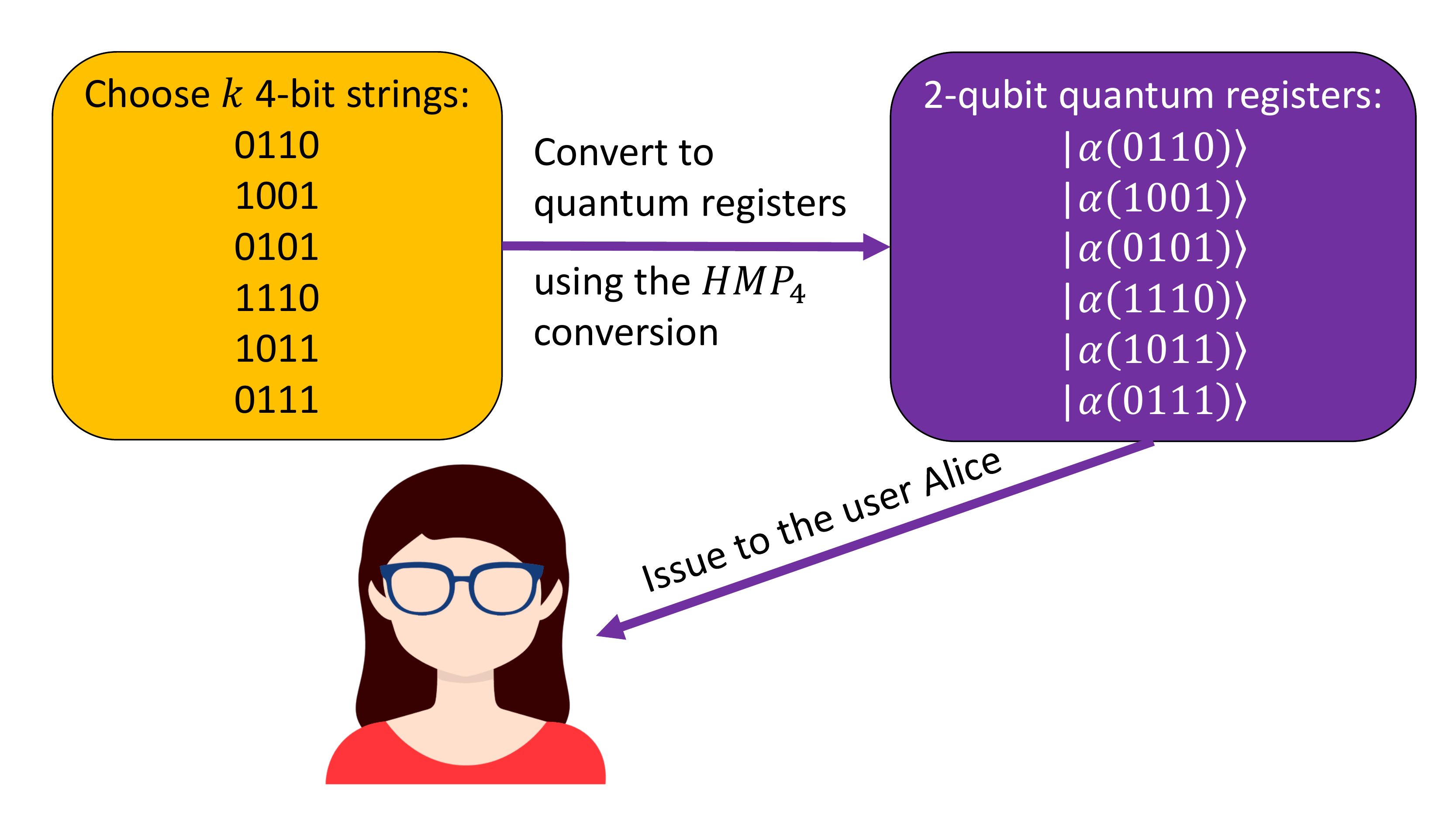}
    \caption{\textbf{Issuing} the quantum MFA token - \textcolor{Fuchsia}{purple: quantum step}, \textcolor{orange}{orange: classical step}.}
    \label{fig:issue}
\end{figure}

The steps for the issuer to create the token with $k$ quantum registers are listed below and and an example is depicted in Figure~\ref{fig:issue}:
\begin{enumerate}
\item Choose an unique identifier for the token, \texttt{tokenID}.
\item Randomly choose $k$ 4-bit strings. We call these $(\mathtt{x}_i)_{i=1..k}$. These bits, along with the token ID are stored in the authentication database. Other details, such as the user that the token is issued to, could also be stored.
\item Convert each $(\mathtt{x}_i)_{i=1..k}$ into a corresponding $\textit{HMP}_4$-state according to Def.~\ref{def:HMP-states} in Section~\ref{sec:background-quantum}.
\end{enumerate}
The $\textit{HMP}_4$-states can now be issued to a user in a token.

\subsection*{Authentication}
In the second phase, the user wishes to proceed with authentication using their quantum token. Depending on the parameters of the scheme, authentication with the same token can be performed multiple times. The user connects to the server, and is issued with a measurement challenge, during which it is decided which quantum memories will be measured and the details of those measurements (e.g. basis). If the user has access to the quantum memories, then the measurements will be performed and the server can validate the results of the measurements against their classical database. Someone impersonating the user will not be able to perform the measurements, and will have to guess the results (possibly based on eavesdropped information), and will fail with high probability. Note that a measurement will collapse the quantum register, and prevent future useful measurements.

\begin{figure}
    \centering
    \includegraphics[width=0.8\linewidth]{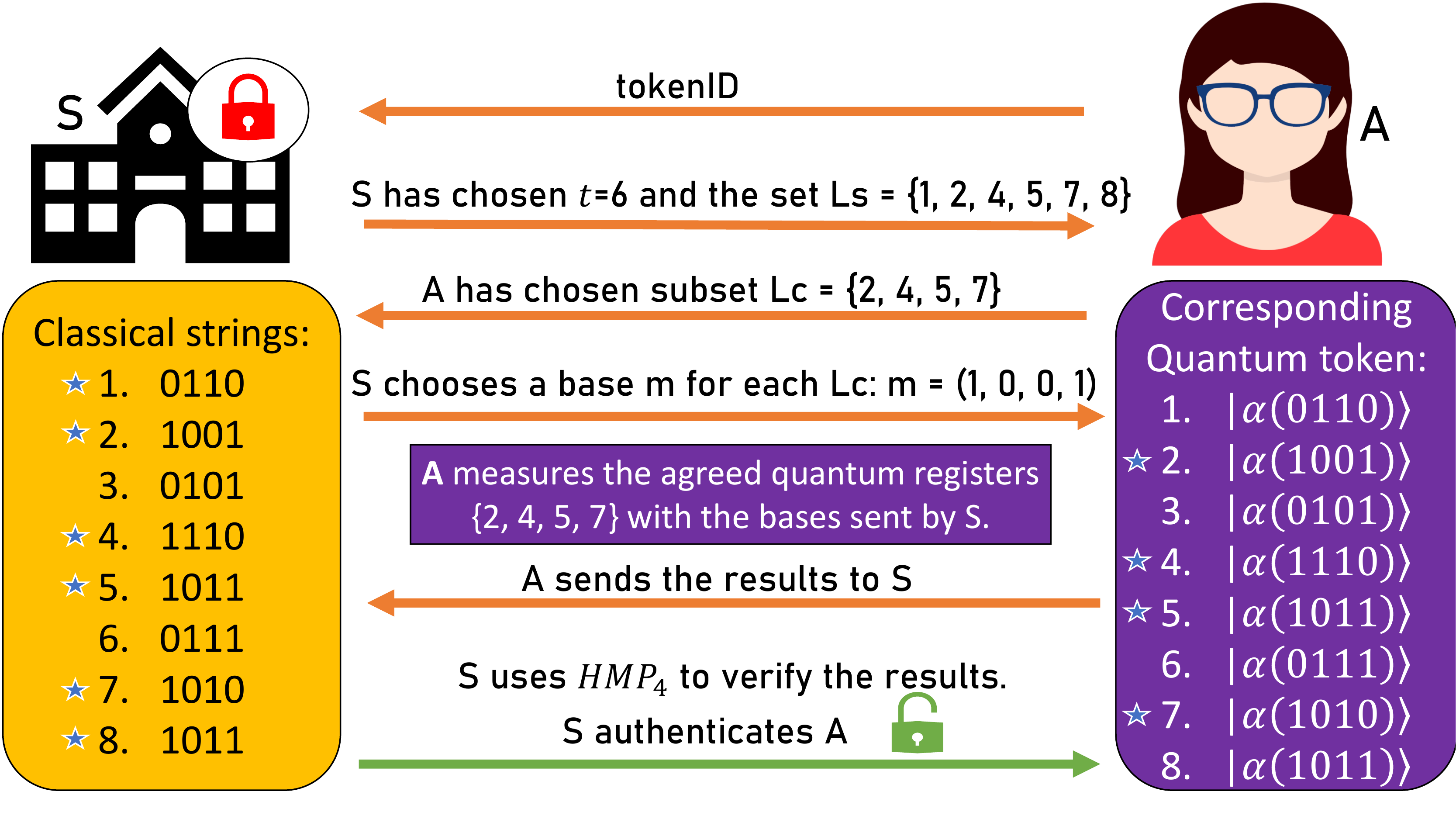}
    \caption{\textbf{Authentication} using the quantum MFA token - \textcolor{Fuchsia}{purple: quantum step}, \textcolor{orange}{orange: classical step}.}
    \label{fig:auth}
\end{figure}

To authenticate, the user must prove they have access to the token, so a challenge is issued. The size of the challenge, $t$, is another parameter of the scheme. For convenience, $t$ should be a multiple of three. The steps for the server and the client are listed below and an example is depicted in Figure~\ref{fig:auth}:
\begin{enumerate}
\item \label{step:tokenID} The user sends the \texttt{tokenID} to the server so that records can be found in the authentication database. This data is static over the token's lifetime.
\item \label{step:set} The server then randomly chooses a set $\mathtt{L_S}$ of $t$ indices from the set $1..k$ of candidate quantum registers to measure.
\item \label{step:subset} Some of these registers may already have been used, so the user randomly chooses a subset $\mathtt{L_C}$ of $\mathtt{L_S}$ of size $2t/3$ of registers that have not yet been used, and sends $\mathtt{L_C}$ to the server. If there is no such subset, then the token has been used up, and a new token should be requested.
\item \label{step:mi} The server then picks random bits $(\mathtt{m}_i)_{i\in\mathtt{L_S}}$ for each of the registers that the client has nominated. These $\mathtt{m}_i$ determine the measurements that the client must perform. The scheme is designed so that the measurement corresponding to $\mathtt{m}_i = 0$ tells us nothing about the result of the measurement for $\mathtt{m}_i = 1$, and vice versa.
\item \label{step:measure} The client performs the measurements on the registers $i \in \mathtt{L_S}$ according to the values $\mathtt{m}_i$, producing two bits $(a_i,b_i)_{i \in \mathtt{L_S}}$ for each measurement. These are sent to the server. The measurements are specified as part of the HMP\textsubscript{4} problem, and dictate the basis used. The user marks the measured registers as used.
\item \label{step:verify} The server verifies that the $(\mathtt{a}_i,\mathtt{b}_i)$ all have the expected values, which can be calculated from $\mathtt{x}_i$ and $\mathtt{m}_i$ using HMP\textsubscript{4}~\cite{bar2004exponential}.
\end{enumerate}

Note that all communication is classical and therefore each client does not require a quantum communication channel to the server in order to authenticate.

\subsection{SASL mechanism}
To demonstrate that this mechanism can be used for authentication in classical systems, we will describe the verification stage, once the user has the quantum token as a Simple Authentication and Security Layer (SASL) mechanism \cite{melnikov2006simple}. In SASL mechanisms, a client (C) is authenticating to a server (S). The server has a list of the IDs of the quantum tokens and also the associated classical strings. The client has the quantum token and an identity that they wish to be authorized as. In SASL terms, this is a client-first protocol, as the client presents information to the server first. The SASL conversation would proceed as follows:
\begin{description}
\item[\texttt{C:}] Authenticate as \texttt{identity} using \texttt{tokenID}.
\item[\texttt{S:}] Measurement challenge is on the set $\mathtt{L_S}$.
\item[\texttt{C:}] I will measure the subset $\mathtt{L_C}$.
\item[\texttt{S:}] HMP\textsubscript{4} challenge measurements are $\mathtt{m}_i$ for $i \in \mathtt{L_C}$.
\item[\texttt{C:}] HMP\textsubscript{4} response is $(\mathtt{a}_i,\mathtt{b}_i)$ for $i \in \mathtt{L_C}$.
\item[\texttt{S:}] Authentication success/failure.
\end{description}

Here, the server would return an authentication success only if all of the following hold: (1) \texttt{tokenID} is in the authentication database (2) \texttt{identity} is permitted authenticate using \texttt{tokenID}; (3) $\mathtt{L_C} \subset \mathtt{L_S}$; (4) the HMP\textsubscript{4} response is given for the same $i \in \mathtt{L_C}$; and (5) the $(\mathtt{a}_i,\mathtt{b}_i)$ values satisfy the relations given by $\mathtt{m}_i$ and $\mathtt{x}_i$ in Definition~\ref{def:HMP}.

\subsection{Informal Security Analysis}

Note, that this protocol can take place in the clear. For each authentication observed, the observer learns the measurements for $2t/3$ registers under the bases $\mathtt{m}_i$. The selection of registers in Step~\ref{step:set} by the server means that an impersonator who has observed an authentication cannot depend on being asked for a subset of registers that they have observed measurements for. Indeed, even if an impersonator has seen multiple measurements, so that every register has been measured, they only observed each register being measured in one of two possible bases specified by $\mathtt{m}_i$. This means that they will know the correct $(\mathtt{a}_i,\mathtt{b}_i)$ values for (on average) half of the $2t/3$ registers, but have to guess the responses for the other half, giving a low probability of success, which reduces exponentially as $t$ increases. The protocol is subject to a MITM attack, though it could be secured via TLS or a similar mechanism.

Theft of the quantum token is, of course, possible. A more interesting question asks,  ``Can the token be cloned?''. Of course, the no cloning theorem means that the quantum registers cannot be cloned by an attacker who steals the token and wants to return it before the theft is noticed. Gavinsky considers the possibility of cloning the quantum token via more complex measurements, and bounds the probability of success. For example, he shows that only  after $e^{\Omega(t^3/k^2)}$ observations cloning might be possible with a probability higher than  $e^{\Omega(t^2/k)}$.

%We should look at the probability of compromise for current 2FA, using a mobile device, and compare to this. We really should compare to an integrated app 2FA setting too as this would have much better security (in particular it would/should have an encrypted commms channel). Maybe we might conclude that offers more security than text based 2FA (not susceptible to spoofing, zero knowledge proof etc), and that it offers at least as good security guarantees as an app based 2FA but at the moment comes with additional costs due to the lack of quantum deployability and re-issuing - though actually a random number generator is pretty intensive.

Note that since each measurement uses up a number of quantum registers this does create an attack: the attacker could convince a user to repeatedly authenticate until enough registers have been measured that authentication becomes impractical. If this happens, the user would have to get a new token issued. 

\subsection{Token Lifetime}

Gavinsky provides various results about how many times this verification procedure can be used. Each verification uses $2t/3$ registers, and he suggests that the token be renewed when $k/4$ registers have been used, to ensure that Step~\ref{step:subset} can be completed with high probability. 

An interesting property of this protocol is that the token's lifetime is limited by the number of uses. As registers are measured, it becomes increasingly less likely that an honest user will be able to successfully complete Step~\ref{step:subset}. Even an attacker who steals the token, who might proceed beyond Step~\ref{step:subset} faces the challenge that the $\mathtt{m}_i$ chosen by the server will not correspond to previous measurements, and will have to resort to guessing, which is analogous to the case of the impersonator above. Gavinsky shows that combinations of quantum measurements can not be used to answer multiple HMP\textsubscript{4} queries with confidence.

This allows us to make a trade-off between the number of registers, $k$, the number of registers used for each authentication, $t$, and the lifetime of the token. It is even possible for the server to choose a $t$ value, depending on the level of authentication required. For example, if authenticating for a low-risk service or if multiple factors are presented, the server could choose a smaller value for $t$ to extend the token lifetime. %XXX Mention step up auth here?

\section{Comparison to state of the art}\label{sec:state-art}
In this section, we will briefly comment on the benefits and disadvantages of the current state-of-the-art mechanisms for multi-factor authentication. We have included One-time passwords (SMS and application based), hardware tokens and the recent work by Sharma et al.~\cite{sharma2020two} on a quantum-based one time password scheme. After detailing each alternative method, we then describe the contributions of our mechanism and the additional benefits and features it offers. 

\paragraph{One time passwords}
The first mechanism we will discuss is one-time passwords. Here, the factor involves providing an additional security code that changes over time. These are commonly either communicated via text/call to a phone or a specific application. Both involve usually a human entering the security code and can be prone to typographic errors, though devices are now providing features to make this easier.

The text/call version requires a phone network connection in addition to an internet connection. The connection to the phone network is treated as a second channel over which to deliver the security code. The text/call version is no longer considered a secure practice for one time passwords~\cite{schneiern-sms}. There are a number of reasons for this, including: (1) SMS and voice calls are not encrypted, so the password could be intercepted and copied; (2) phone company employees can be fooled into transferring a phone number to the attacker's SIM card, meaning the security codes get sent to the attacker.

The application-based version of one-time passwords usually requires each user to download additional software. Time-based one time passwords (TOTP) is a common form. This involves hashing an agreed secret (sometimes sent using a QR code) with a time stamp, in order to generate a time-dependent code. Provided that the client's application and the verifier know the initial secret and the time, they will both be able to generate the same time-based code and the client can provide this to the server for authentication. Sometimes this mechanism is implemented using a hardware token with a clock and a pre-loaded secret. Each time-based code is agreed to be valid for a precisely defined time interval, usually 1 to 15 minutes. However, if the initial secret is stolen, then whoever has this initial secret will also be able to generate an arbitrarily large number of valid codes.

Both one-time password mechanisms, and our quantum mechanism, depend on generation of secret random numbers (i.e., the random code in the voice/text mechanism, the initial secret TOTP and the $\mathtt{x}_i$ in the quantum mechanism).
The one-time password mechanisms usually require human interaction, which is practical because the schemes just require the entry of a simple code. The quantum mechanism is more complicated, and we believe it would only be practical to implement as part of an automated protocol (such as SASL).
Relative to the voice/text mechanism, the quantum mechanism does not require an additional channel after the token is issued, and so it is more like the application-based mechanism.
Relative to the application-based mechanism, the quantum scheme cannot have the secret stolen from the user's token, as any quantum registers that are copied/measured, will collapse. Note, that an attacker that has access to the server-side database (the $\mathtt{x}_i$) can effectively impersonate someone with the token.

\paragraph{Hardware tokens}

A second common mechanism for multi-factor authentication is the issuing of a hardware device (such as a device with display or a specifically designed USB/NFC/BLE device) to each user.
We mentioned one form of hardware token above, where TOTP is used with a dedicated hardware token with a pre-loaded secret.
A second common example of this is a device supporting FIDO2/WebAuthn, such as a YubiKey, though these mechanisms can also exist on a mobile device with a hardware security module.

Like TOTP and our quantum mechanism, these devices have a registration phase. Here, the device produces a public/private key pair that can be used for signing challenges. The public key is stored on the server at registration, and the private key is stored in a way that is only accessible to the hardware device. Once registration is complete, the device can be used for authentication where the server presents a challenge and information to allow the identification of the correct key. The challenge is then signed by the device (following confirmation by the user through a gesture, such as pushing a button or providing a biometric). The signature can be verified by the server using the public key stored by the server at registration.

A crucial advantage of schemes based on public key cryptography is that the server does not store the private key and so compromise of the server does not permit an attacker to impersonate users in the future. This is an advantage over both the quantum scheme and schemes such as TOTP. These devices are also often advertised as phishing resistant, as under normal operation the device will not reveal the private key. However, there is a possibility to extract the private keys from the hardware device, and, though highly challenging, this has been demonstrated~\cite{lomne2021side}. There is also no limit on how many times a public key can be used, in comparison with the quantum scheme. We also note that many public-key signature schemes are vulnerable to attack by quantum computing. A quantum token should be resistant to such attacks. We note that several features common to hardware tokens, such as a push button, biometric verification or incorporation into a mobile device could be used as part of the implementation of the quantum MFA scheme.

\paragraph{Biometric based Quantum One time passwords}
Sharma and Nene~\cite{sharma2020two} suggest a quantum two-factor authentication mechanism which combines a biometric code, $B_C$, and a quantum challenge. They propose three variations of the protocol and for each one the user registers their biometric with the server in the form of a shared secret code. 

The first form of the protocol is an introductory mechanism. The server generates a series of two entangled qubits and measures one of the qubits in each pair to get a code, $C_1$. Instead of storing the code, the server stores the corresponding entangled qubit. The code, $C_1$ is sent to the Client via SMS, then the client computes $C_1 \oplus B_C$ and sends the result back to the server. By measuring the second entangled qubit and XORing this with the stored biometric code for the user, the server can verify that the user did, in fact, receive the code to their device. This mechanism involves quantum memory and quantum measurement capabilities for the server. The benefit over a classic SMS-based one-time password scheme are that the qubit stored at the server will collapse if an attacker attempts to duplicate it and that the quantum entanglement is being used for the random generation of the code. 

The second form of the protocol involves the server sending the qubit to the client and then the client making the measurement. This has the advantage of securing the transmission as the qubits would collapse if observed. It means that quantum communication is now required for each authentication verification and the client and the server will require quantum measurement capabilities. In our mechanism, the authentication after registration involves only classical communication, but similar to this protocol we do require the client to have quantum measurement capabilities. 

The third form of the protocol suggested by Sharma and Nene~\cite{sharma2020two} involves bi-directional quantum communication and quantum gate capabilities for the user. The server will also still need quantum measurement capabilities. The difference in this mechanism is that the client will transform their biometric code to a Pauli matrix before XORing it with the qubit that the server sends. The result is a qubit in a different Bell state, which is communicated back to the server. No classical channel is used in this case, therefore a Man-in-the-middle (MITM) attack is not possible as an eavesdropper will collapse the qubits' wave functions. The server can then use the corresponding entangled qubits to verify that the user has returned the correct values.

It is also worth noting that the inclusion of the biometric in all these schemes brings its own problems. A biometric scanner does not always give a clear reading and the code generated, even with error-correcting, can vary each time~\cite{fhloinn2006biometric}. When this code is sent to the server it must match the code they have on file. The biometric code can also be stolen in transmission at registration, or recorded on the device and replayed by an attacker.  

%These schemes give an attacker a 50:50 change of guessing a qubit measurement correctly. In our scheme, by introducing measurement with respect to different bases, the attacker has a 25\% chance of making a correct guess per qubit. This could be included in Sharma and Nene~\cite{sharma2020two} scheme by agreeing a random basis as part of the protocol, for example using Diffie-Hellman to agree a basis per qubit. 

The advantage of our scheme over these schemes lies primarily in the use of a token, which can last for multiple authentications and the fact that the server does not need to store qubits for each client's authentication. Furthermore, the authentication procedures only involve classical communication, and because it is a form of a zero-knowledge protocol, this classical channel does not necessarily need to be encrypted, though the simple inclusion of a TLS connection will give protection against a MITM attack.

%It is phishing resistant, a body pretending to be the server does not get all the info needed to authenticate. They will cause the user to destroy registers unnecessarily, but only serves to nullify the value of the information the attacker has gathered. 

%A MITM attack does has some effect in our mechanism, the attacker will be able to record a qubit's measurement with respect to one of the two possible basis. If they pretend to be the user, they would now be able to guess the correct responses with a 50\% chance instead of 25\%. However, notice that a 50\% guessing chance per qubits still gives it the same security level as the Sharma and Nene scheme~\cite{sharma2020two}.

\subsection{Attack Susceptibility Summary}
In Table~\ref{tab:compare}, we summarize a comparison of the different multi-factor authentication mechanisms. For each attack type, we indicate whether the mechanism is susceptible to the attack or not. Notice that we do not include MITM attacks, as every mechanism will be susceptible to a single session being compromised by MITM, but can avoid it by using a TLS connection. Biometric-QOTP~3 uses only quantum communication channels for authentication, so cannot, and does not need to, secure it via TLS. However, this introduces the challenge that the quantum authentication will need to be tied to the future classical communication in some way in order to make the authentication effective.

\begin{table}
    \begin{center}
    \begin{tabular}{|l||C{0.7cm}|C{0.7cm}|C{0.7cm}|C{0.7cm}|C{0.7cm}|c|}\cline{2-7}
         \multicolumn{1}{c||}{} &\rot{\shortstack[l]{Phishing}} & \rot{Replay} & \rot{\shortstack[l]{Eavesdropper\\analysis}} & \rot{Keyboard logging }  & \rot{Clone-able}&  \rot{\shortstack[l]{Channels for\\authentication}}\\\hline
        Our QMFA &\N&  \N &\N & \N & \N & $C$ \\
        SMS-based OTP & \N &  \N & \Y & \Y  & \Y &$C, T$\\
        App-based OTP & \N & \N & \Y & \Y  & \Y&  $C$ \\
        FIDO2 (Public-key token) & \N &  \N & \N & \N  & \Y &  $C$\\
        Biometric-QOTP~1& \N  & \N & \Y & \Y & \Y&  $C, T$\\
        Biometric-QOTP~2& \N & \N & \N & \Y  & \Y&  $Q_1, C$\\
        Biometric-QOTP~3& \N & \N & \N & \Y  & \Y &  $Q_2$\\\hline
    \end{tabular}
    \end{center}
    {\footnotesize
    Vulnerability: (\N) Not susceptible, (\Y) Susceptible. \\  Channel type: (C)~Classical, (T)~Telephone/mobile, ($Q_1$)~Quantum~(1-way), ($Q_2$)~Quantum~(2-way).}
    \vspace{1ex}
    \caption{Attacks each multi-factor authentication mechanism is susceptible to. 
    }
    \label{tab:compare}
\end{table}

The other attacks are defined as follows.
\begin{description}
\item[Phishing:] An attacker masquerading as a website tries to trick the user into revealing their secret to them. None of the MFA mechanisms above are subject to this attack.
\item[Replay:] An attacker who records the user's authentication response should not be able to directly replay it in a later authentication in order to gain access. Again, none of the state-of-the-art mechanisms are vulnerable to this.
\item[Eavesdropper analysis:] An eavesdropper can listen and record communication between the server and the user. They can then use/analyse this data offline and potentially use it to log in on a different session.
%If TLS is not used, then the secret shared code (based on the biometric, QR code, etc) could be uncovered.
Note that we assume that TLS cannot be used to secure a SMS based system when communication is via the traditional telecommunications network. %  if this correct for eavesdropping???
\item[Keyboard logging:] These attacks involve malware (or malicious hardware) on the user's device which can record user inputs. Both biometrics and typed codes are subject to this attack. Note that once the biometric is stolen, the attacker can successfully authenticate as the user in future for all the Biometric-QOTP mechanisms.
\item[Clone-able:] This considers whether an attacker can duplicate the authenticator so that they have the power to authenticate. A classic example is sim-jacking, where an attacker convinces a phone company to pass a phone number, or SIM, over to them instead of the true owner~\cite{vaughan2020phone}. The app-based OTP is clone-able, as, if the initial secret is discovered, an attacker can login indefinitely as the user. Similarly, if the biometric code is cloned then the holder of the code can login as the user for future authentications. A FIDO public-key based token is clone-able under extreme circumstances~\cite{lomne2021side}.
\end{description}

The last column indicates the channels that are needed during the authentication process. $Q_1$ refers to the need for a 1-way quantum communication channel. For example, from the server to the user. $Q_2$ is the requirement for 2-way quantum communication, so that the user is also able to use a quantum channel for their response. $T$ indicates the need for a telephone communication channel and $C$ indicates a classical internet connection.

\section{Discussion}\label{sec:disc}

The quantum scheme we have proposed here has several interesting features, including features common to modern schemes, such as being resistant to phishing and eavesdropping attacks, while also including having a limited number of uses and quantum-level resistance to cloning. It also avoids quantum storage at the server and the need for a quantum channel when authenticating.

\paragraph{Token Distribution} Challenges in distributing hardware tokens with pre-shared secrets could apply to this scheme. For, say, employees who have a remote/in person working arrangement updating the registers in their quantum token might be straight forward, and would allow periodic in-person validation. However, for purely remote users, things could be more challenging, with tokens either being delivered by post (and subject to theft) or via some quantum network.

\paragraph{Implementation Challenges}
Of course, the current state of quantum computing, and particularly for this scheme, quantum memories, means a true hardware implementation of this scheme is not currently possible. We have implemented Gavinsky’s quantum coin scheme within a quantum communications simulator~\cite{murray2020implementing}, and confirmed that the main implementation challenges relate to the availability of quantum hardware. This would allow us, in future work, to implement the scheme and study the noise and errors that occur within it and test possible potential attacks. Quantum technology is evolving at a rapid pace, and having trialed and tested potential mechanisms for when implementation is possible is important for reliable cryptography and security. 

\paragraph{Variable Strength Authentication} As we noted, a server could potentially choose different challenge sizes, $t$, when authenticating a user. These could be chosen for different levels of security. For example, a user connecting from a commonly-used IP address could be challenged with a small $t$, whereas a user connecting from an unexpected country might be challenged with a larger $t$ value. An additional challenge might also be used for \emph{step-up authentication} when a user wants to access higher-risk systems, for example password change or access to personal data. This could be particularly interesting, given that each quantum register read provides additional confidence to the server, while smaller $t$ values would extend the token's lifetime.

\section{Conclusion}\label{sec:conc}
In this paper, we have introduced a quantum multi-factor authentication protocol. It is based on the hidden-matching quantum-complexity problem. It offers step-up graded authentication for users via a quantum token. Advantages over classical schemes include the automatic protection against duplication and eavesdropping which is inherent to quantum bits. It has the ability to offer high and low levels of assurance with the same token, based on current needs and actions of the user. Also, it is not susceptible to known quantum computing attacks unlike many classical asymmetric authentication mechanisms.
%Some key benefits of this mechanism in comparison to quantum equivalents include: a secure communication channel is not a hard-requirement for the authentication, only classical communication is required for each authentication (rather than quantum communication), the verifier need also store classical bits rather than quantum registers, it allows variations in the degrees of security requested which can be adapted by the verifier.
We have compared it to a number of classical MFA schemes, highlighting benefits and challenges.
Overall, it is a promising mechanism that, given the existence of quantum memories, could prove valuable. 
\bibliographystyle{splncs04}
\bibliography{lib}

\begin{thebibliography}{10}
\providecommand{\url}[1]{\texttt{#1}}
\providecommand{\urlprefix}{URL }
\providecommand{\doi}[1]{https://doi.org/#1}

\bibitem{bar2004exponential}
Bar-Yossef, Z., Jayram, T.S., Kerenidis, I.: Exponential separation of quantum
  and classical one-way communication complexity. In: Proc.
  36\textsuperscript{th} ACM symposium on Theory of computing. pp. 128--137
  (2004)

\bibitem{barnum1999quantum}
Barnum, H.N.: Quantum secure identification using entanglement and catalysis.
  arXiv preprint quant-ph/9910072  (1999)

\bibitem{bennett1983quantum}
Bennett, C.H., Brassard, G., Breidbart, S., Wiesner, S.: Quantum cryptography,
  or unforgeable subway tokens. In: Advances in Cryptology. pp. 267--275.
  Springer (1983)

\bibitem{duvsek1999quantum}
Du{\v{s}}ek, M., Haderka, O., Hendrych, M., My{\v{s}}ka, R.: Quantum
  identification system. Physical Review A  \textbf{60}(1), ~149 (1999)

\bibitem{fhloinn2006biometric}
Fhloinn, E.N.: Biometric Retrieval of Cryptographic Keys. Ph.D. thesis, Trinity
  College Dublin (2006)

\bibitem{gavinsky2012quantum}
Gavinsky, D.: Quantum money with classical verification. In: 2012 IEEE
  27\textsuperscript{th} Conference on Computational Complexity. pp. 42--52.
  {IEEE} (2012)

\bibitem{ghilen2013quantum}
Ghilen, A., Azizi, M., Bouallegue, R., Belmabrouk, H.: Quantum authentication
  based on entangled states. In: Proc. of World Cong. on Multimedia and
  Computer Science. pp. 75--78 (2013)

\bibitem{gonzalez2021attack}
Gonz{\'a}lez-Guill{\'e}n, C.E., Gonz{\'a}lez~Vasco, M.I., Johnson, F.,
  P{\'e}rez~del Pozo, {\'A}.L.: An attack on {Z}awadzki’s quantum
  authentication scheme. Entropy  \textbf{23}(4), ~389 (2021)

\bibitem{grover1996algorithms}
Grover, L.K.: A fast quantum mechanical algorithm for database search. In:
  Proc. 28\textsuperscript{th} ACM Symposium on Theory of Computing. p.
  212–219. STOC ’96, Association for Computing Machinery, New York, NY, USA
  (1996). \doi{10.1145/237814.237866},
  \url{https://doi.org/10.1145/237814.237866}

\bibitem{ho2017quantum}
Hong, C.h., Heo, J., Jang, J.G., Kwon, D.: Quantum identity authentication with
  single photon. Quantum Information Processing  \textbf{16}(10),  1--20 (2017)

\bibitem{kuhn2003hybrid}
Kuhn, D.R.: A hybrid authentication protocol using quantum entanglement and
  symmetric cryptography. arXiv preprint quant-ph/0301150  (2003)

\bibitem{kunnemann2012yubisecure}
K{\"u}nnemann, R., Steel, G.: {YubiSecure}? {F}ormal security analysis results
  for the {Yubikey} and {YubiHSM}. In: International Workshop on Security and
  Trust Management. pp. 257--272. Springer (2012)

\bibitem{li2004quantum}
Li, X., Barnum, H.: Quantum authentication using entangled states.
  International Journal of Foundations of Computer Science  \textbf{15}(04),
  609--617 (2004)

\bibitem{li2006quantum}
Li, X., Zhang, D.: Quantum information authentication using entangled states.
  In: International Conference on Digital Telecommunications (ICDT'06). pp.
  64--64. IEEE (2006)

\bibitem{lomne2021side}
Lomn{\'e}, V., Roche, T.: A side journey to {T}itan. IACR Cryptol. ePrint Arch.
   \textbf{2021}, ~28 (2021)

\bibitem{mavroeidis2018impact}
Mavroeidis, V., Vishi, K., Zych, M.D., J{\o}sang, A.: The impact of quantum
  computing on present cryptography. arXiv preprint arXiv:1804.00200  (2018)

\bibitem{melnikov2006simple}
Melnikov, A., Zeilenga, K.: Simple authentication and security layer ({SASL}).
  IETF RFC  \textbf{4422} (2006)

\bibitem{murray2020implementing}
Murray, H., Horgan, J., Santos, J.F., Malone, D., Siljak, H.: Implementing a
  quantum coin scheme. In: 2020 31st Irish Signals and Systems Conference
  (ISSC). pp.~1--7. IEEE (2020)

\bibitem{m2011totp}
M’Raihi, D., Machani, S., Pei, M., Rydell, J.: {TOTP}: Time-based one-time
  password algorithm. IETF RFC  \textbf{6238} (2011)

\bibitem{schneiern-sms}
Schneier, B.: {NIST is No Longer Recommending Two-Factor Authentication Using
  SMS}.
  \url{https://www.schneier.com/blog/archives/2016/08/nist_is_no_long.html}
  (Aug 2016)

\bibitem{sharma2020two}
Sharma, M.K., Nene, M.J.: Two-factor authentication using biometric based
  quantum operations. Security and Privacy  \textbf{3}(3), ~e102 (2020)

\bibitem{shor1994algorithms}
Shor, P.W.: Algorithms for quantum computation: discrete logarithms and
  factoring. In: Proc. 35\textsuperscript{th} Symposium on Foundations of
  Computer Science. pp. 124--134. {IEEE} (1994)

\bibitem{vaughan2020phone}
Vaughan, A.: Phone number theft through sim-jacking is on the rise in the uk.
  New scientist (3264), ~10 (2020)

\bibitem{wooters1982quantum}
Wooters, W., Zurek, W.: Quantum no-cloning theorem. Nature  \textbf{299}, ~802
  (1982)

\bibitem{zawadzki2019quantum}
Zawadzki, P.: Quantum identity authentication without entanglement. Quantum
  Information Processing  \textbf{18}(1),  1--12 (2019)

\bibitem{zeng2000quantum}
Zeng, G., Guo, G.: Quantum authentication protocol. arXiv preprint
  quant-ph/0001046  (2000)

\end{thebibliography}

\end{document}